# THE ROTATION OF GALAXIES
# AS A CONSEQUENCE OF
# A ROTATION OF THE UNIVERSE

**E Chaliasos**
365 Thebes Street
GR-12241 Aegaleo
Greece

*Abstract*

Assuming a suitable rotation of the Universe as a whole, we can attribute the rotation of galaxies to it. Using Lagrangian Mechanics we then can find the equations governing the rotation of a galaxy. We find in this way the azimuthal equation, and we find its first integral. Then we derive from it the law of differential rotation of galaxies, and we plot it. We also present the resulting flat rotation curve, which is found to resemble remarkably the observed rotation curves.

**1. Introduction**

Starting from the observation of the stricking similarity between the spiral galaxies (see 1st image) and the cyclones on Earth (see 2nd image), the author has made the hypothesis that this similarity is due to similar causes. But we all know that the spiral structure of cyclones is a consequence of the rotation of Earth. Thus, the hypothesis implies that a hypothetical rotation of the Universe as a whole may cause the rotation of galaxies. The conspecious spiral structure of many galaxies will then be attributed to the rotation of the Universe.

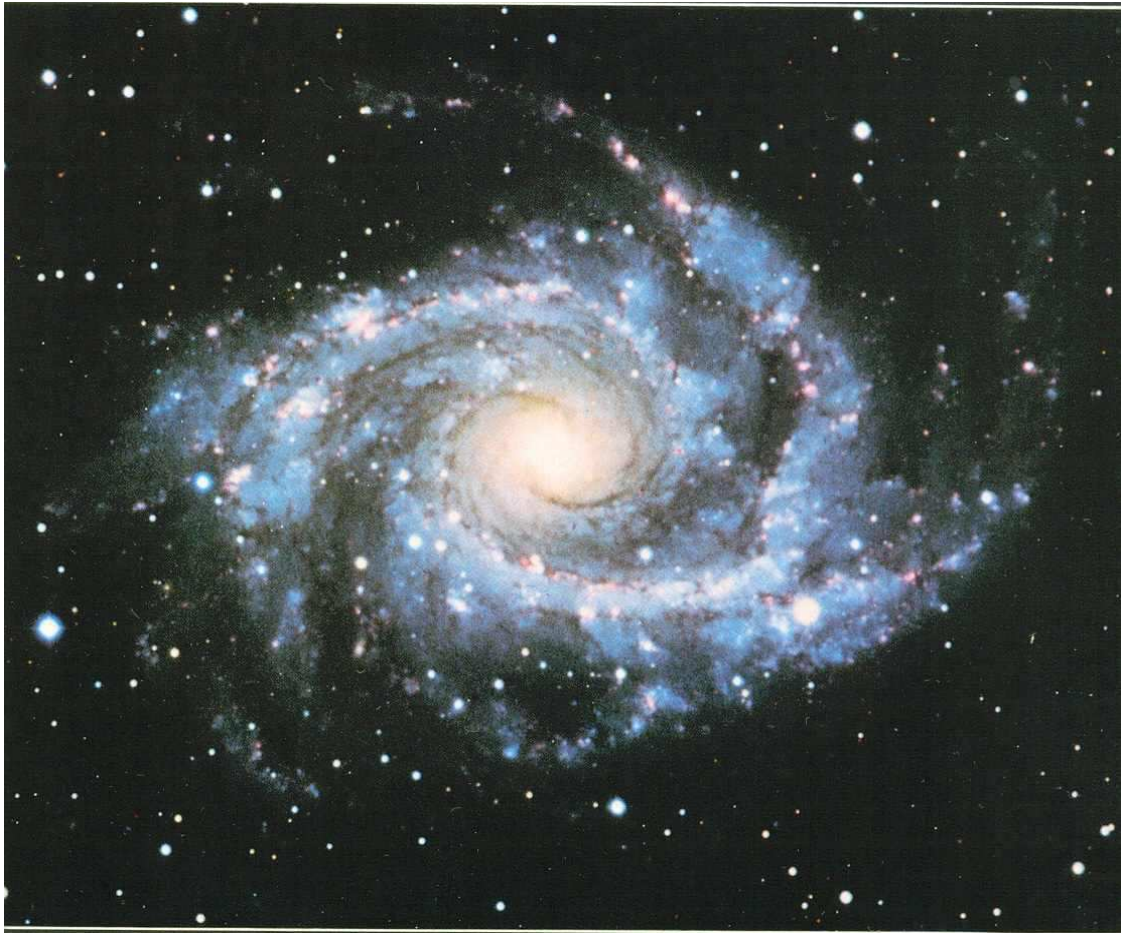

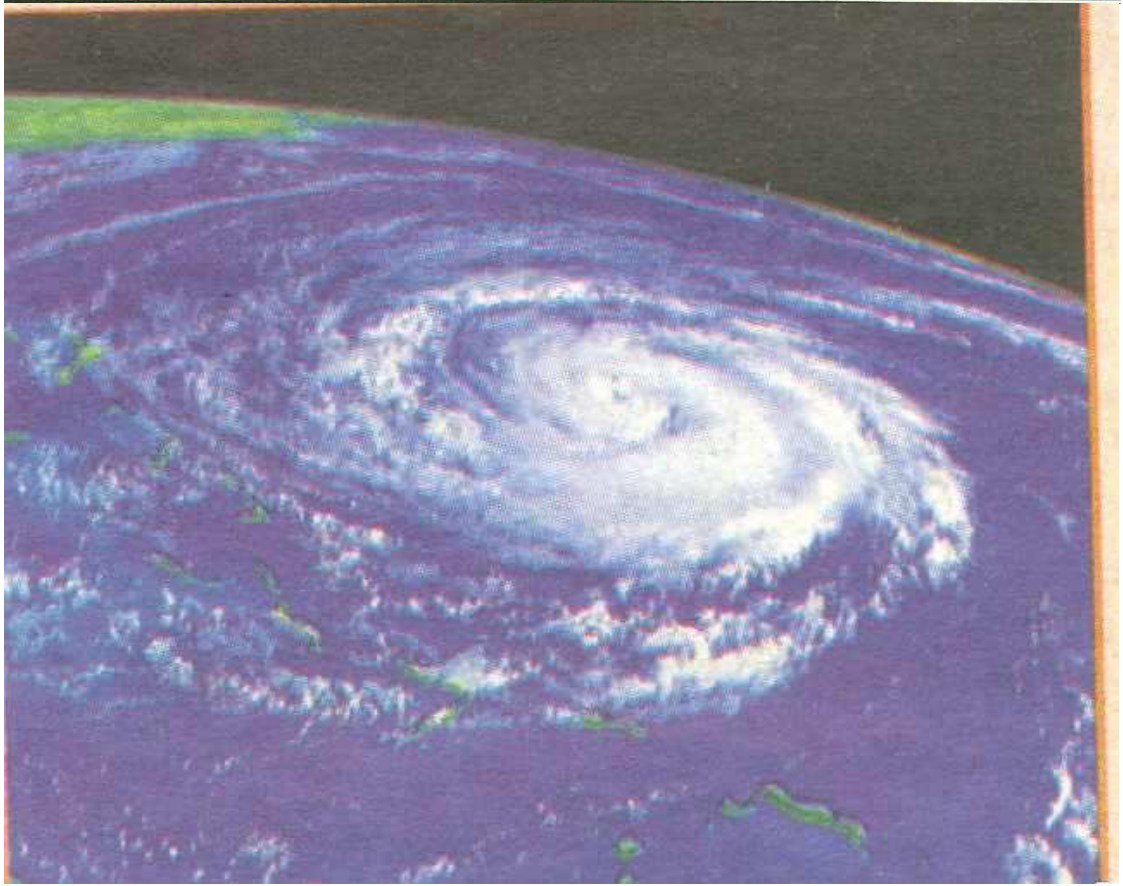

But what kind of rotation could be attributed to the Universe as a whole? The author has answered this question by postulating that our three dimensional physical space can be embedded in a fictitious four dimensional space in which it executes the rotation. Note that this 4-space has nothing to do with the four dimensional space-time. He studied theoretically such a rotation using the General Theory of Relativity, and exposed the results of this theoretical research in his paper [1].

The way was right by solving the Einstein equations starting from an appropriate form of the metric. A summary of the results could be found in [2]. The main ones were that the scale factor varies as exp(Gt) & the two angular velocities of the general rotation of the Universe involved vary as exp(-3Gt), where t is the world time and G a constant which in turn coincides with the Hubble constant. In addition it was found that the matter density was constant in time which implied a steady state Universe, and that the equation of state was right p+e = 0, where p the pressure and e the energy density of the cosmic fluid.

But the General Theory of Relativity was found that was not necessarily the only way to study the rotation of the Universe. We can actually study this rotation by merely Newtonian Mechanics [3]. In the present paper we will study the impacts of this rotation of the Universe on the structure and internal motions of galaxies. And we will specifically focus on the explanation of the observed flat rotation curves of galaxies, by finding theoretically by the use of Classical Mechanics the law of differential rotation of galaxies.

## 2. The rotation

Suppose that the 3-space $(x_1, x_2, x_3)$ is rotated about the axis $x_4$, with angular velocity $\omega$. It is evident that the axis $x_4$ has only one point in common with the 3-space $(x_1, x_2, x_3)$, the origin. The rotation, thus, of the 3-space about the $x_4$-axis is equivalent to a rotation of $(x_1, x_2, x_3)$ about a fixed point, the origin O. But such a rotation involves three idependent rotations. This is equivalent to the rotation of $(x_1, x_2, x_3)$ about a 3-dimensional axis, which however is *not fixed,* and the rotation of $(x_1, x_2, x_3)$ about this *instanteneous* axis of rotation has the magnitude of angular velocity $\omega$. In fact, then, the three independent rotations can be taken as the projections of $\omega$ on the three coordinate planes, namely $\omega_1$ on $x_2Ox_3$, $\omega_2$ on $x_3Ox_1$ and $\omega_3$ on the $x_1Ox_2$.

The relation of $\omega_1$, $\omega_2$, $\omega_3$ to $\omega$ can be found as follows. Suppose that in time dt the 3-space is rotated around the instantaneous axis through an angle $d\theta$, while the direction of the axis is given by the three direction cosines $\alpha$, $\beta$, $\gamma$. If in this infinitesimal time interval the corresponding angle of rotation $d\theta$ has the projections $d\theta_{yz}$, $d\theta_{zx}$, $d\theta_{xy}$ on the corresponding coordinate planes, then we will have the relations [4]

$$\left. \begin{array}{l} \delta\theta_{yz} = \alpha d\theta \\ d\theta_{zx} = \beta d\theta \\ d\theta_{xy} = \gamma d\theta \end{array} \right\} \quad (1)$$

or, dividing by dt,

$$\left. \begin{array}{l} \omega_1 = \alpha\omega \\ \omega_2 = \beta\omega \\ \omega_3 = \gamma\omega \end{array} \right\} \quad (2)$$

but, since

$$\alpha^2 + \beta^2 + \gamma^2 = 1, \quad (3)$$

it is clear that

$$\omega_1^2 + \omega_2^2 + \omega_3^2 = \omega^2. \quad (4)$$

Now, if an elementary coordinate volume dV, at the point P, has the edges $ds_1 = rd\theta_{yz}$, $ds_2 = rd\theta_{zx}$, $ds_3 = rd\theta_{xy}$, where r is the distance of P from O, then

$$dV/dt^3 = r^3 \omega_1 \omega_2 \omega_3. \quad (5)$$

Assuming that this is an invariant during the rotation, it follows that

$$\omega_1 \omega_2 \omega_3 \sim 1/r^3. \quad (6)$$

If we start with a surface element instead of a volume one, then similarly we obtain

$$\omega_2 \omega_3 \sim 1/r^2 \quad (\omega_1 = 0), \quad \text{etc.} \quad (7)$$

and, if we start with a line element,

$$\omega_1 \sim 1/r \quad (\omega_2 = \omega_3 = 0) \quad \text{etc.} \quad (8)$$

See the *Appendix* for a clarification of this point.

Since, now, the 3-space is rotated, it is no longer isotropic, with the result that the law of conservation of angular momentum is not strictly valid. We can only suppose that it is approximately valid, if we assume that the magnitude of the rotation is small. Then the plane of rotation of a galaxy will be approximately unaltered by the rotation. Evidently the axis of rotation is then fixed for a concrete galaxy, and the rotation of this galaxy on its plane will be $\omega \sim 1/r$.[*] Thus, concerning a galaxy we can write its angular velocity of rotation as $\omega/r$, with $\omega$ a <u>constant</u>.

### 3. The angular equation of motion and its first integral

To find the equation we will use Lagrangian Mechanics, in cylindrical coordinates, as these coordinates are appropriate for our problem. Thus, we can write for the kinetic energy T of a star moving about in the galaxy

$$T = (1/2)m(\dot{r}^2 + r^2\dot{\phi}^2) \tag{9}$$

[5], where m is the mass of the star and r, $\phi$ its polar coordinates (we disregard the z-axis). The Lagrange equation will be

$$\frac{d}{dt}\left(\frac{\partial T}{\partial \dot{\phi}}\right) - \frac{\partial T}{\partial \phi} = Q_\phi \tag{10}$$

[6], where $Q_\phi$ the angular generalized force, that is the torque. We have

$$\frac{\partial T}{\partial \dot{\phi}} = r^2\dot{\phi} \quad \& \quad \frac{\partial T}{\partial \phi} = 0. \tag{11}$$

Concerning $Q_\phi$, we have to observe that the only tangential forces acting on the star are the Coriolis tangential component $2(\omega/r)r'$, and the one due to the variation of $(\omega/r)$. The latter is $\mathbf{r} \times (\mathbf{\omega}/r)'$, with magnitude $r\omega(d/dr)(1/r)r'$. The corresponding torques are then $2\omega r'$ and $-\omega r'$. Thus, the Lagrange equation is

$$(r^2\ddot{\phi}) = \omega\dot{r}. \tag{12}$$

A first integral of this equation is

$$r^2\dot{\phi} = \omega r - K^2, \tag{13}$$

where $-K^2$ is a constant of integration.[**] Thus the angular velocity of the star $\Omega$ (=$\phi'$) follows the law

$$\Omega = \frac{\omega}{r} - \frac{K^2}{r^2}. \tag{14}$$

The tangential linear velocity of the star V (=$r\phi'$) follows therefore the law

$$V = \omega - \frac{K^2}{r}. \tag{15}$$

---

[*] It is clear that the planes of rotation of the various galaxies will be radomly distributed.

[**] The constant of integration has to be taken negative and (absolutely) very small in order to have agreement with the observation.

### 4. The rotation curves of galaxies

Differenciating (14) we find

$$\Omega' = -\frac{\omega}{r^2} + \frac{2K^2}{r^3}. \tag{16}$$

Equating this to zero we obtain

$$r = \frac{2K^2}{\omega}. \tag{17}$$

Differenciating again (16) results in

$$\Omega'' = \frac{2\omega}{r^3} - \frac{6K^2}{r^4}. \tag{18}$$

Equating this to zero leads to

$$r = \frac{3K^2}{\omega}. \tag{19}$$

Thus the curve $\Omega(r)$ presents a maximum at $2K^2/\omega$ and a turning point at $3K^2/\omega$ (see Fig. 3)

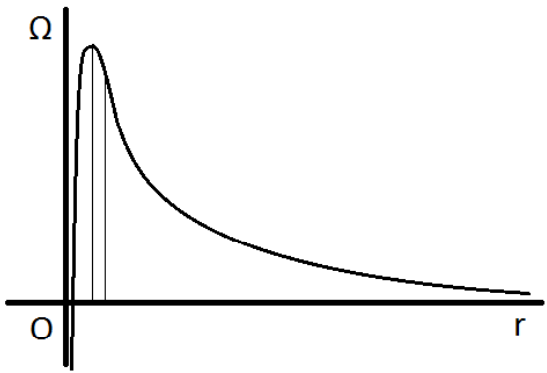

Fig. 3

Differenciating (15) we find

$$V' = \frac{K^2}{r^2} > 0. \tag{20}$$

Differenciating it again we obtain

$$V'' = -\frac{2K^2}{r^3} < 0. \tag{21}$$

Thus the curve V(r) has the form presented in Fig. 4.

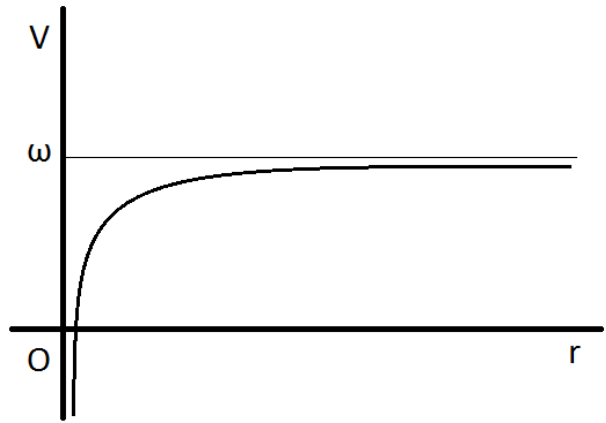

Fig. 4

Observe the amazing agreement with the observations [7, p. 600]. Note that we can find ω from measurements on the curve of Fig. 4. We must find the same ω for all galaxies. This will give a measurement of the angular velocity of the Universe as a whole!

**Appendix**

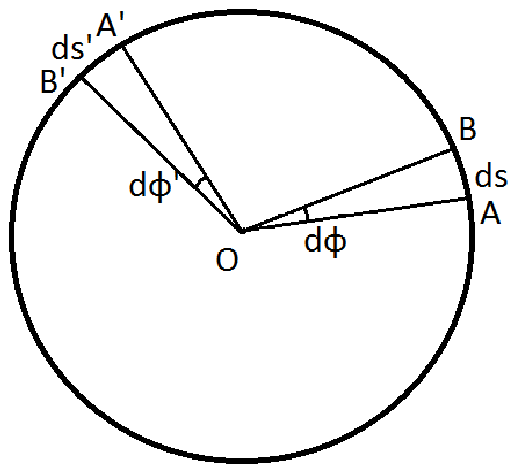

Fig. 5

Let us take into consideration the rotation on a plane about O with angular velocity ω. If ω depends only on the radius r, and if we consider r fixed at the moment, the motion of the points with constant r will be on a circle (see Fig. 5). Let A→B & A′→B′ in time dt. We have AA′→BB′ by the rotation in time dt. We assume that the rotation leaves the arc lengths invariant on the circle. Then AA′ = BB′, and therefore AA′−BA′ = BB′−BA′, that is ds = ds′. But ds = rdϕ & ds′ = rdϕ′. Therefore the angle of rotation in time dt is inversely proportional to r, and we can write ω ≡ dϕ/dt ~ 1/r. We can then write ω as ω/r, with ω a constant, independent of r.

## REFERENCES

1. Chaliasos, E. (2006): «The rotating and accelerating Universe», arXiv.org/abs/astro-ph/0601659
2. Chaliasos, E. (2009): «The rotating and accelerating Universe», ASPCS, vol. 424, Edited by Tsinganos, K. *et al,* San Francisco
3. Chaliasos, E. (2011): «Cosmological results from a Newtonian rotation of the Universe», ABI Chronicles, Raleight (N.C., U.S.A.)
4. Misner, C.W., Thorne, K.S., Wheeler, J.A. (1973): «Gravitation», Freeman & Co., San Francisco
5. Landau, L.D. & Lifshitz, E.M. (1976): «Mechanics» (3rd edn), vol. 1 of «Course of Theoretical Physics», Pergamon, Oxford
6. Goldstein, H. (1968): «Classical Mechanics», Addison-Wesley, Reading
7. *Binney, J. & Tremaine, S. (1987): «Galactic Dynamics», Princeton, N.J.*